\documentclass[conference]{IEEEtran}
\IEEEoverridecommandlockouts
\usepackage{cite}
\usepackage{amsmath,amssymb,amsfonts}
\usepackage{textcomp}
\usepackage{xcolor}

\usepackage{stfloats}
\usepackage{url}
\usepackage{verbatim}
\usepackage{graphicx}
\usepackage{multibib}
\usepackage{subfigure}
\usepackage{url}

\usepackage[linesnumbered,ruled]{algorithm2e}  

\usepackage{color}
\usepackage{amssymb} 
\usepackage{multirow}
\usepackage{multicol}
\usepackage{float} 
\usepackage{booktabs}
\usepackage{arydshln}
\usepackage{listings}
\definecolor{green}{RGB}{146,208,080}

\def\BibTeX{{\rm B\kern-.05em{\sc i\kern-.025em b}\kern-.08em
    T\kern-.1667em\lower.7ex\hbox{E}\kern-.125emX}}
\begin{document}

\title{MPipeMoE: Memory Efficient MoE for Pre-trained Models with Adaptive Pipeline Parallelism 
}

\author{

\IEEEauthorblockN{Zheng Zhang}
\IEEEauthorblockA{\textit{School of Computer Science} \\
\textit{WuHan University} \\
zzhang3031@whu.edu.cn}
\and

\IEEEauthorblockN{Donglin Yang}
\IEEEauthorblockA{\textit{Nvidia Corp.} \\
dongliny@nvidia.com}
\and

\IEEEauthorblockN{Yaqi Xia}
\IEEEauthorblockA{\textit{School of Computer Science} \\  \textit{WuHan University} \\
yaqixia@whu.edu.cn}
\and

\IEEEauthorblockN{Liang Ding}
\IEEEauthorblockA{\textit{JD Explore Academy} \\ \textit{JD.com Inc.} \\
liangding.liam@gmail.com}
\and

\quad  \quad \quad \quad \quad  \quad \quad 
\IEEEauthorblockN{Dacheng Tao}
\IEEEauthorblockA{\quad \quad \quad \quad  \quad \quad  \quad  \textit{JD Explore Academy} \\ 
\quad \quad \quad \quad \quad  \quad \quad \textit{ JD.com Inc.} \\
 \quad \quad \quad \quad \quad  \quad \quad  dacheng.tao@gmail.com}
\and

\IEEEauthorblockN{Xiaobo Zhou}
\IEEEauthorblockA{\textit{University of Macau} \\
waynexzhou@um.edu.mo}
\and

\IEEEauthorblockN{Dazhao Cheng}
\IEEEauthorblockA{\textit{School of Computer Science} \\ \textit{WuHan University} \\
dcheng@whu.edu.cn}
\and
}

\maketitle

\begin{abstract}
Recently, Mixture-of-Experts (MoE) has become one of the most popular techniques to scale pre-trained models to extraordinarily large sizes. Dynamic activation of experts allows for conditional computation, increasing the number of parameters of neural networks, which is critical for absorbing the vast amounts of knowledge available in many deep learning areas. However, despite the existing system and algorithm optimizations, there are significant challenges to be tackled when it comes to the inefficiencies of communication and memory consumption.

In this paper, we present the design and implementation of MPipeMoE, a high-performance library that accelerates MoE training with adaptive and memory-efficient pipeline parallelism. Inspired by that the MoE training procedure can be divided into multiple independent sub-stages, we design adaptive pipeline parallelism with an online algorithm to configure the granularity of the pipelining. Further, we analyze the memory footprint breakdown of MoE training and identify that activations and temporary buffers are the primary contributors to the overall memory footprint. Toward memory efficiency, we propose memory reusing strategies to reduce memory requirements by eliminating memory redundancies, and develop an adaptive selection component to determine the optimal strategy that considers both hardware capacities and model characteristics at runtime. We implement MPipeMoE upon PyTorch and evaluate it with common MoE models in a physical cluster consisting of 8 NVIDIA DGX A100 servers. Compared with the state-of-art approach, MPipeMoE achieves up to 2.8$\times$ speedup and reduces memory footprint by up to 47\% in training large models.

\end{abstract}


\section{Introduction}

Scaling up the model size of neural networks is one of the promising ways to improving model accuracy in a wide range of applications~\cite{b1, b3, antman, whale, axonn, wang2020efficient}. For example, in natural language processing (NLP), large pre-trained language models~\cite{bert, roberta, t5, gpt3} have been shown effective in many domains such as language understanding~\cite{t5}, sequence generating~\cite{zhang2022bliss, zhong2022e2s2} and cross-lingual downstream transfer~\cite{b8, zhong2022toward}. Recently, Mixture-of-Experts (MoE) has been adopted to scale neural networks to an extreme size without introducing a proportional increase in computational cost~\cite{moe, gshard, switch, he2022cherry}. The MoE architecture consists of many sub-models called~\textit{experts}. It employs a trainable gating network to  intelligently forward the input token to specific experts. The sparse combination of experts makes it practical to save much computation capacity and improve model accuracy compared to dense models with the same computation resources. There are popular MoE-based models in recent years such as Google's Switch Transformer\cite{switch} and Meta's BASE Layer\cite{baselayer}.

\begin{figure}[t] 
    \centering
    \scalebox{0.60}{
    \includegraphics[width=.8\textwidth,trim=0 0 0 0,clip]{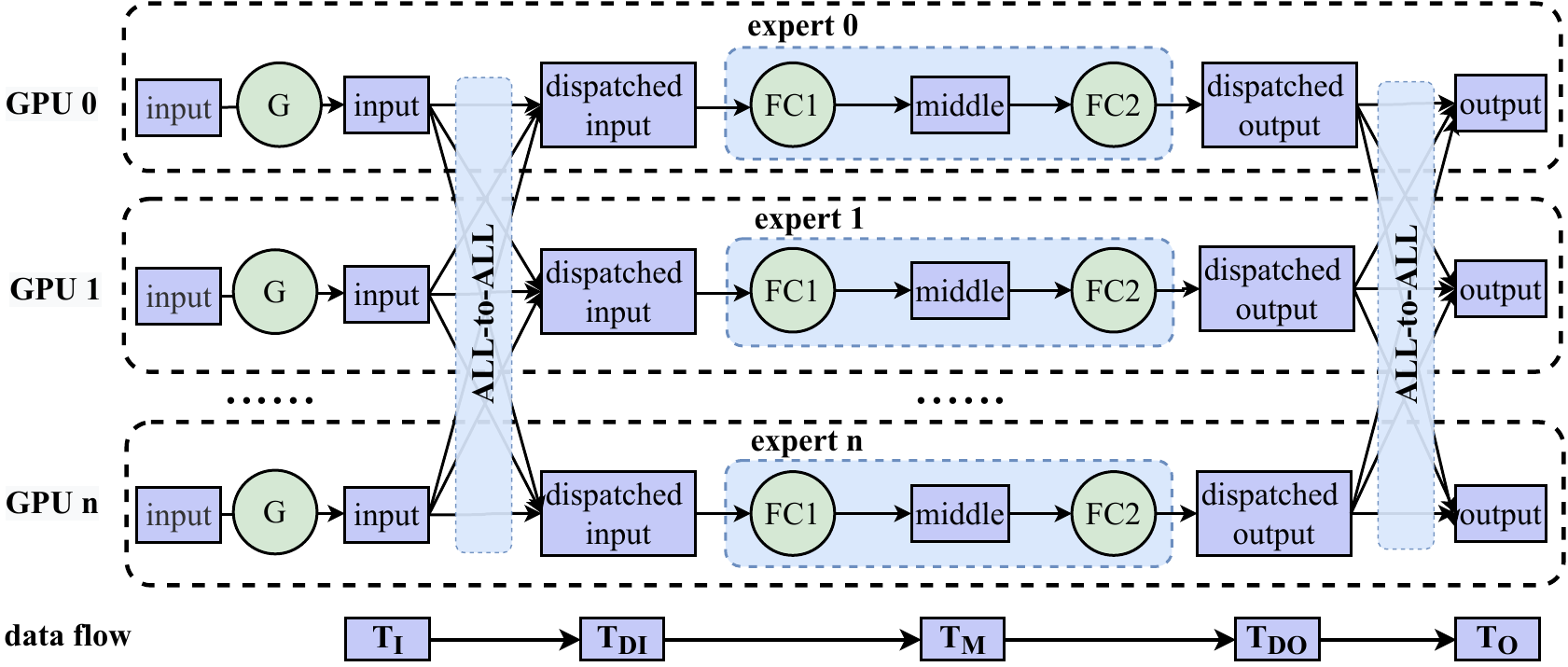}
    }
    \caption{The illustration of expert parallelism of MoE and its data flow. The green circles represent sub-modules of the MoE layer, and the purple rectangles represent activation tensors of MoE training. 
    For simplicity, We take $T_I, T_{DI}, T_{M}, T_{DO}, T_{O}$ at the bottom of the figure as abbreviations of \textit{input, dispatched input, middle, dispatched output, output} tensors, which are in purple color.}
    \label{fig:moe-flow}
\end{figure}

For training a MoE model, different experts are distributed across a large number of GPU servers. The training process requires All-to-All communication primitive operations to dispatch tokens to the desired experts and collect them after processing. This procedure is called expert parallelism~\cite{switch}, which is shown in Figure~\ref{fig:moe-flow}. In a distributed fashion, the main performance bottleneck comes from the communication phase. It is reported in literature~\cite{gating-drop} that a variant of MoE without All-to-All can achieve a relative improvement of communication cost for more than 90\% in extreme cases. 
Besides, when scaling up model at extra-scale, the limited size of GPU DRAM has been a major challenge for researchers to explore deeper and wider neural networks.

Training a giant MoE model at the trillion scale requires tremendous hardware resources. For example, training a model consisting of 600 billion parameters in GShard~\cite{gshard} takes up to 96 hours on a cluster equipped with 2,048 TPUs. 
There are system and algorithm optimizations that tackle the intrinsic inefficiency of All-to-All synchronous communication in MoE~\cite{gating-drop,gshard,ds-moe,fastermoe}. 
For example, the work~\cite{gating-drop} proposed a gating dropout algorithm to reduce the traffic of communication. Recently, FasterMoE~\cite{fastermoe} adopts pipeline parallelism to alleviate the overhead of communication with expert shadowing. It can achieve significant speedup upon the existing systems in training large MoE models. However, the granularity of pipelining is pre-defined and it is fixed throughout the training. In practice, the dynamic nature of communication demands for adaptive pipeline parallelism, because coarse-grained pipelining is sub-optimal in taking advantage of parallelism while very fine-grained pipelining incurs significant overhead because of frequent kernel launches and GPU under-utilization. Furthermore, the existing approaches ignore memory efficiency in MoE training, which is however key to scaling up the model to extra-scale.


In this paper, we propose to address the inefficiency of communication and memory usage of MoE training in a holistic manner. First, to alleviate the overhead of communication, we analyze the system behaviors of communication and computation for the MoE architecture and design adaptive pipeline parallelism for MoE~\cite{wang2019scalable}, which partitions a batch of tokens into several micro-batches and overlaps the execution of computation and communication.  Different from FasterMoE~\cite{fastermoe}, we partition tokens in a more effective manner and propose an adaptive configuration algorithm to search for the optimal pipeline granularity.


Furthermore, we examine the memory footprint of MoE training, which mainly comes from three components: i) \textit{model states of experts}, which include parameters, optimizer states and gradients; 
ii) \textit{activations}, which need to be stored/stashed in the forward pass so that they can be used later in the backward pass; iii) \textit{temporary buffers}, which store the gradients of activations during the backward pass that are discarded as soon as they are used. Among the three components, activations are the primary contributor to the memory footprint when the batch size is increased. As shown in Figure~\ref{fig:moe-flow}, expert parallelism~\cite{switch} is designed to scale up the model size by distributing experts across devices evenly. Similarly to Zero Redundancy Optimizer~\cite{zero, zcodem3}, it partitions parameters, optimizer states, and gradients of the model across devices, alleviating the memory footprint of model states in MoE. However, the memory footprints of activations and temporary buffers have the potential for further reduction.


We propose to reduce the memory footprint of activations and temporary buffers by sharing the same buffer for different partitions of tensors. Specifically, the memory of tensors $T_{DI}, T_{M}, T_{DO}$ can be vastly reduced from $m$ to $\frac{m}{n}$, in which $m$ refers to the memory requirement and $n$ is the number of partitions that determines the granularity of pipelining. 
But a new challenge is introduced, as activations are overwritten when different partitions request the same memory address. To deal with this problem, we resort to re-computation/communication~\cite{sublinear-mem} and CPU offloading~\cite{vdnn, buddy} for recovering activations in the backward pass. When the re-computation is enabled, the cost of computation can be overlapped with that of the communication, and vice versa. In addition, leveraging that modern GPUs support overlapping computations and data transfers over PCIe, we can offload data to the CPU in the forward pass and prefetch the data into GPUs accordingly. Specifically, $T_{DI}$ can be obtained by either communication or CPU offloading while $T_{M}$ can be obtained by either re-computation or CPU offloading. 
We establish a performance model to configure the ideal strategy at runtime. 


In summary, we make the following contributions.
\begin{itemize}
\item We design adaptive pipeline parallelism for MoE by partitioning a batch of tokens into several micro-batches and overlapping the execution of computation and communication to improve the utilization of GPUs and network bandwidth. We present an online search algorithm to configure the optimal pipeline granularity.
\item We analyze the memory footprint breakdown of MoE and find that activations and temporary buffers are the primary contributors to the memory footprint. With the pipeline parallelism, we propose to reduce the memory footprint of activations and temporary buffers by sharing the same memory buffer for different partitions. 
\item We tackle the problem that activations are overwritten when different partitions request the same memory space. We leverage re-computation/re-communication and CPU offloading for recovering activations in the backward pass based on performance modeling.
\item We implement and integrate the proposed techniques into a library for MoE training, namely MPipeMoE. Experimental results show that MPipeMoE can achieve up to 47\% memory footprint reduction and 2.8$\times$ speedup over the state-of-the-art system FasterMoE.
\end{itemize}


The rest of this paper is organized as follows. Section~\ref{sec:preliminaries} gives background and motivations for distributed training of MoE models. Sections~\ref{sec:designs} and \ref{sec:impl} describe the system design and implementation of MPipeMoE, respectively. 
Section~\ref{sec:exp} presents the experimental setup and evaluation results. 
Section~\ref{sec:related} reviews related works. 
Section~\ref{sec:conclude} concludes the paper.



\section{Background and Motivation} \label{sec:preliminaries}

\subsection{Mixture of Experts~(MoE)}
The transformer architecture was introduced to the NLP community due to its superior performance in sequence-to-sequence tasks, such as neural machine translation. A transformer model consists of a few blocks, each of which is formed by the self-attention, cross-attention, and Feed-Forward-Network (FFN) modules. Ever since, transformer-based models become the top performers in various NLP tasks, such as BERT~\cite{bert}, RoBERTa~\cite{roberta}, and GPT-3~\cite{gpt3}. Scaling up the model size results in a significant increase in computational cost for both training and inference. For example, it takes 168 days to train a GPT-3 model with 178 billion parameters using 256 NVIDIA A100 GPUs~\cite{Narayanan:2021}.

MoE provides an efficient solution to reducing the cost of training extra-scale models, which incurs only sub-linear compute costs concerning the model size by sparsely activating a subset of the model parameters for given inputs. For example, the cost of training the Switch Transformer~\cite{switch} with 1.6 trillion parameters are indeed less than the computation budget required to train a dense model with 10 billion parameters. The core component of these MoE models~\cite{moe, zcodem3, switch} is the MoE layer, which replaces the FFN sub-layer in the original dense transformers.

\textbf{Expert Parallelism for MoE}. To train a giant MoE model, expert parallelism\cite{switch} is widely applied to reduce the memory footprint by distributing different experts across devices. As shown in Figure~\ref{fig:moe-flow}, a gating network determines the destination device of each token, which is followed by All-to-All communication. After the dispatch All-to-All, each device executes the local expert, which is an FFN layer consisting of two linear layers and one activation function. Then, the second All-to-All communication is conducted to send the processed tokens back to the devices to which these tokens belong. 

\textbf{Inefficient Synchronous Communication}. Each expert requires All-to-All communication to send/receive tokens to/from other devices. The communication phase becomes one of the most time-consuming factors in training MoE models~\cite{gating-drop, fastermoe}. The All-to-All and expert process procedures are synchronous operations, which are blocked for waiting for the desired data.



\subsection{Memory Footprint of MoE}
\subsubsection{Where did all the memory go} We first analyze the full spectrum of the memory footprint, including model states, activations, and temporary buffers.

\textbf{Model States}.
Model states are one of the main contributors to memory consumption during training, which includes parameters, gradients, and optimizer states~\cite{zero}. 

\textbf{Activations}.
Activations are the intermediate tensors in forward computing, accounting for a significant amount of memory usage~\cite{sublinear-mem}, especially for the large batch size. 

\textbf{Temporary Buffers}.
Temporary buffers are used to store intermediate results for a very short period, which are not required for future computation, i.e., the backward pass. 

\begin{table}[t]
\caption{Notations used in memory usage formulation.}
\centering
\scalebox{0.88}{
\begin{tabular}{clcl} 
\toprule
 Notation & Definition & Notation & Definition \\
 \midrule
M & model dimension & B & the batch size of tokens \\
H & hidden dimension  & E & the total number of experts\\
n & the number of partitions  & N & the number of nodes \\  
\bottomrule
\end{tabular}}
\label{tab:notations}
\end{table}

\begin{figure}[htbp]
\centering
\scalebox{0.3}{
\centerline{\includegraphics{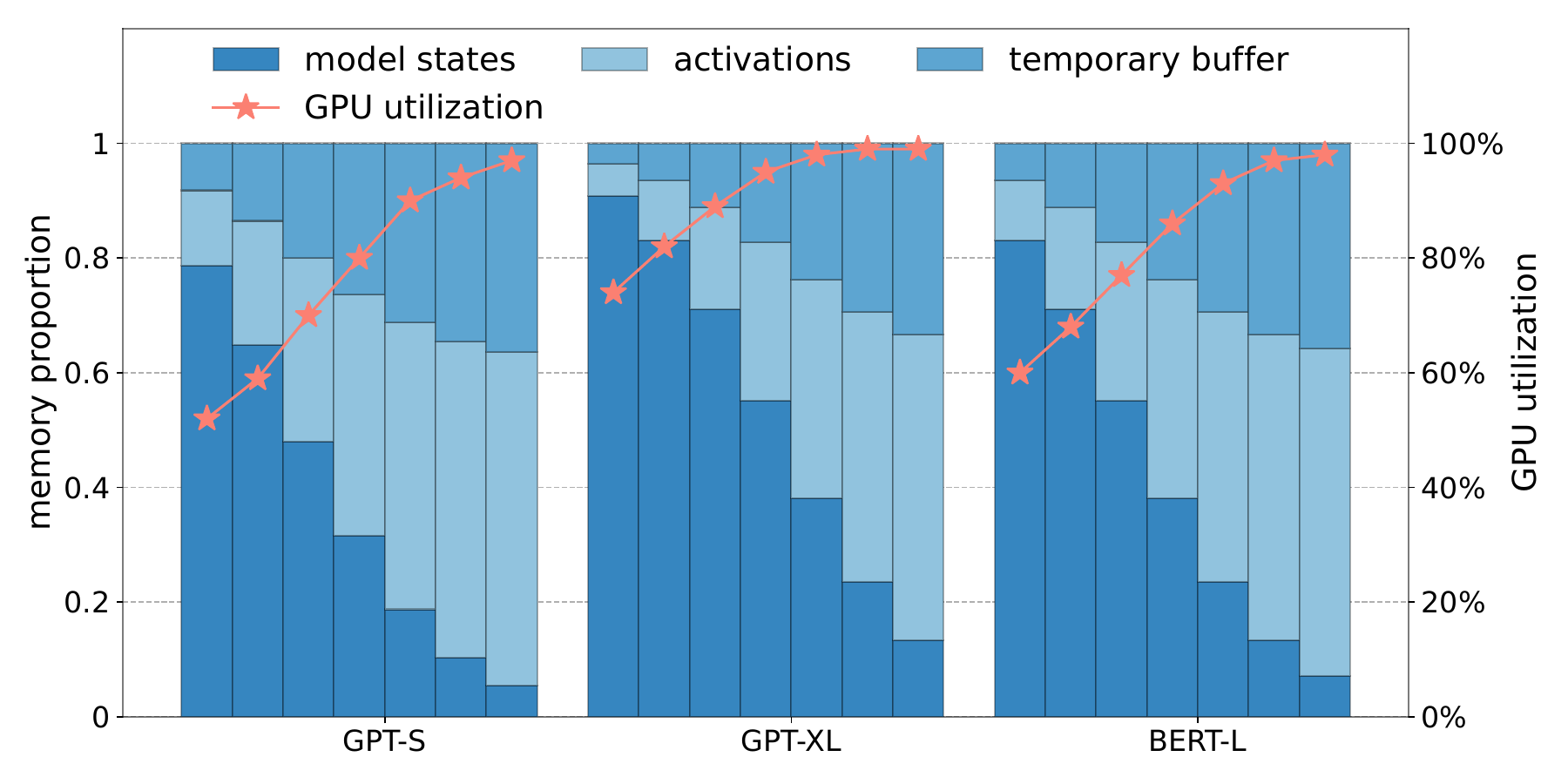}}
}
\caption{Breakdown of memory footprint ratio within model states, activations, and temporary buffers. The experiments are conducted on three different MoE layers with various batch sizes of tokens ranging from 256 to 16k with exponential factor 2.
}
\label{fig:mem-breakdown}
\end{figure}

\subsubsection{Formulation of Memory Footprint of MoE} 
To analyze the memory footprint of MoE, we demonstrate the detailed dataflow of the communication and expert computation, which is shown in Figure~\ref{fig:moe-flow}. Starting with the input tensor $T_I$, the All-to-All stage slices and dispatches the tensor across devices, which is referred to as $T_{DI}$. Every expert takes $T_{DI}$ as the input and outputs tensors  $T_M$ and $T_{DO}$ after two linear layers, i.e., FFNs, in sequence. The activation function is omitted since in-place operations can be applied here. Finally, tensor $T_O$ is obtained by the collective operations on slices of $T_{DO}$.

The memory footprint of model states, activation, and temporary buffers are denoted as $\mathcal{M}_{ms}$, $\mathcal{M}_{act}$, and $\mathcal{M}_{buf}$, respectively.  We summarize other notations in Table~\ref{tab:notations}. The structure of an MoE layer consists of a gating network and an expert. 
As formulated in Equation~\eqref{eq:mem-ms}, $E* M$ equals the number of parameters in the gating network and $2* H* M$ equals that of an expert. Besides, Adam~\cite{adam} is chosen as the default optimizer, requiring an additional memory footprint for momentum and variance. As a result, it takes 4 times the memory of parameters for storing model states, including parameters, gradients, momentum, and variance.


The memory footprint of activations is summarized in Equation~\ref{eq:mem-act}, where the shape of tensors $T_I, T_{Di}, T_{Do}, T_O$ is $(B,M)$ and the shape of tensor $T_{M}$ is $(B, H)$. For simplicity, we do not consider small tensors such as the routing data of the gating network, because their sizes are one to two orders of magnitude smaller than other activation tensors.

In the backward pass, the GPU device is required to allocate temporary buffers to store the gradients of activations which will be discarded as soon as they are used. When operations are executed in sequence, only two adjacent tensors are required to be cached in the device. The formulation of memory footprint is presented in Equation~\ref{eq:mem-bf-seq}, which is the peak requirement of temporary buffers. 
\begin{gather}
\mathcal{M}_{ms} = 4*(E* M+2* H* M)	\label{eq:mem-ms}	\\
\mathcal{M}_{act} = 4* B* M+B* H 	\label{eq:mem-act}	\\
\mathcal{M}_{buf} = B* M+B* H \label{eq:mem-bf-seq}  
\end{gather}

To visualize the memory consumption of the three discussed data types, we plot the ratio of memory footprint in different MoE settings, which are shown in Figure~\ref{fig:mem-breakdown}. It can be seen that activations and temporary buffers account for the major portions of the memory footprint with the increasing number of tokens. We also monitor the GPU utilization for the experiment. We observe that a small batch size leads to GPU under-utilization, especially for the MoE layer in GPT-S. As a result, it is necessary to increase the batch size for higher GPU utilization. Based on the above observations, we motivate the need to reduce the memory footprint of activation tensors and temporary buffers to train the model with the large batch size.

\subsection{Feasibility of Parallelism}
\label{sec:speed-profile}

\begin{figure}[t] 
    \centering
    \scalebox{0.35}{
    \includegraphics[width=.75\textwidth,trim=0 10 0 5,clip]{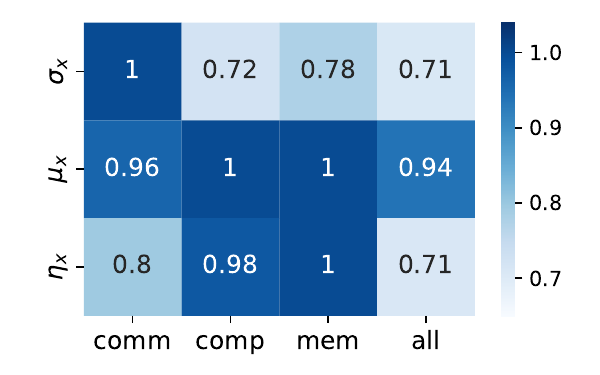}
    }
    \caption{The interference between different operations. The values in the grid represent the relative speed influenced by operations \textit{GeMM computation, communication} and \textit{memory copy}.
    }
    \label{fig:interference}
\end{figure}

The speed of the communication, computation, and memory copy is denoted as $W_{comp}$, $W_{comm}$, and $W_{mem}$, respectively. Ideally, three types of operations do not affect each other when they are being executed in parallel because they request individual hardware resources in principle.
However, in a real environment, there exists resource competition when executing multiple operations in parallel CUDA streams. For example, the communication and memory copy race for memory bandwidth. Performance slowdown incurs if running multiple NVIDIA Collective Communication Library (NCCL) kernels concurrently with computation kernels on the same device.
To quantify the degree of slowdown, we define the actual speed of communication, computation, and memory copy as $\mu_{x}W_{comp}$, $\sigma_{x}W_{comm}$, and $\eta_{x}W_{mem}$, in which $\mu_{x}$, $\sigma_x$, and $\eta_{x}$ represent their corresponding slowdown factors, respectively. The interference stream, i.e., $x$, can be any type of streams such as $comm$, $comp$, and $mem$. Specifically, $all$ is regarded as the case that all three types of CUDA streams are executed in parallel.
The values of $\mu, \sigma$, and $\eta$ indicate the feasibility of parallelism. For example, to take the advantage of overlapping between communication and computation, $\mu_{comm}$ and $\sigma_{comp}$ are required to be greater than $0.5$, otherwise the execution time of communication or computation would exceed the original end-to-end time, leading to deterioration
of the end-to-end performance.

To better understand the interference between operations, we run a micro benchmark in our cluster and measure the actual speed of communication, computation, and memory copy in different situations. Results are demonstrated in Figure~\ref{fig:interference}, from which we can learn that:
\begin{itemize}
    \item Slowdown is introduced in communication if we execute computation with communication in parallel. However, it is feasible to overlap communication and computation as we can make sure that $\mu_{comm}, \sigma_{comp}$ are larger than $0.5$.
    \item Computation is slightly influenced by other operations, which is negligible in terms of end-to-end performance. As a result, we set $\sigma=1$ by default in this paper.
    \item There exists a performance slowdown when communication and memory copy streams are executed in parallel, which is because of bandwidth competition.
\end{itemize}

The observations and analysis above motivate us to design adaptive pipeline parallelism with memory efficiency.

\section{System Design} \label{sec:designs}

\subsection{Overview}
We present the system design of MPipeMoE. First, we design adaptive pipeline parallelism and design an online pipeline granularity configuration algorithm to determine the optimal granularity for accelerating MoE training. Then, we propose the memory reusing component and build a performance model to select the optimal reusing strategy at runtime to reduce the memory footprint.


\begin{figure}[htbp]
\centering
\subfigure[Pipeline parallelism in GPipe.] {
 \label{fig:gpipe}     
\includegraphics[width=0.8\columnwidth]{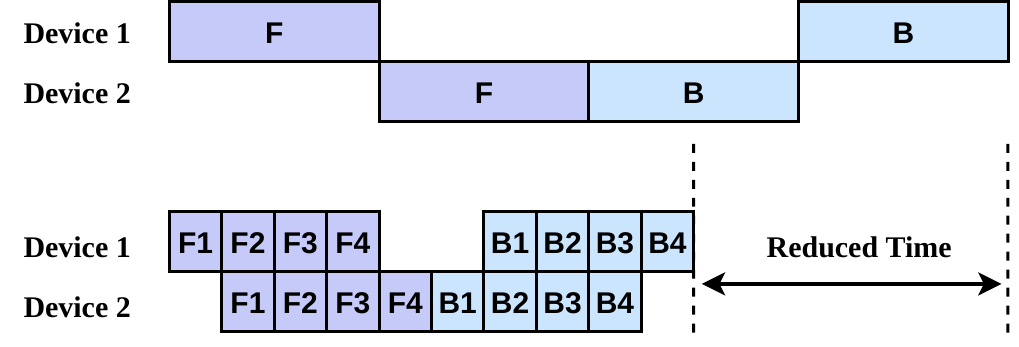}  
}     
\subfigure[The proposed pipeline parallelism.] { 
\label{fig:mpipe}     
\includegraphics[width=0.8\columnwidth]{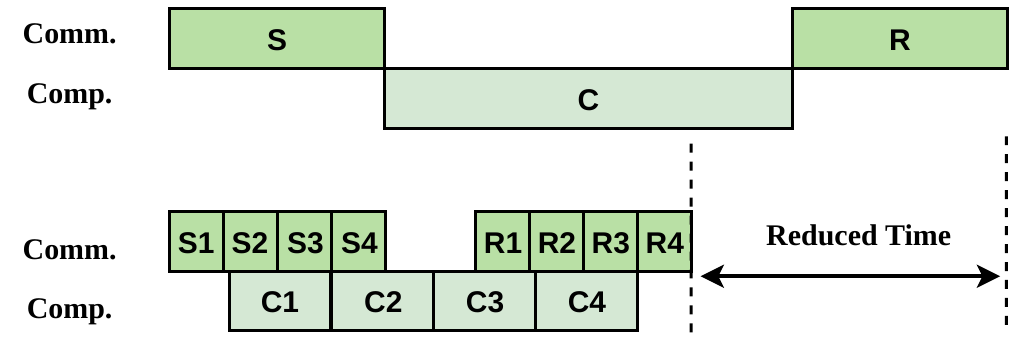}     
}  
\caption{
The illustration of GPipe and micro-batch pipeline parallelism in MPipeMoE. (a) $F$ and $B$ represent forward pass and backward, respectively. (b) $S$, $C$, and $R$ represent the first All-to-All, computation of experts, and the second All-to-All. The serial number of every block represents the index of the micro-batch partition.
}
\label{fig:seq-pipe}
\end{figure}

\subsection{Micro-batch Pipelining} \label{sec:design-pipeline}

 As stated previously, the All-to-All operation is the performance bottleneck to scaling out the training of MoE models. Pipeline parallelism, which is firstly introduced in GPipe~\cite{gpipe},
 can reduce the overhead of communication by overlapping the computation and communication. As is shown in Figure~\ref{fig:gpipe}, layers of the model are partitioned into multiple stages, which are mapped to separate devices for performing computation. To deal with the severe under-utilization caused by the sequential dependency of the neural network, GPipe divides the input mini-batch into smaller micro-batches, allowing different accelerators to work on different micro-batches simultaneously. Inspired by GPipe, the micro-batch parallelism can also be applied to the MoE layers to achieve end-to-end speedup. Note that pipeline is not a new idea, 
 enabling adaptive pipelining for MoE requires online scheduling and the insight of computation separation because of the complex dependencies. The unique contribution of this paper lies in  tackling these specific challenges in a holistic manner.


\textbf{Micro-batch pipelining for MoE}.
As shown at the top of Figure~\ref{fig:mpipe}, only one mini-batch is active for computation or communication in the traditional expert parallelism. In this setup, computation and communication are `idle' most time. With this in mind, we split a mini-batch of tokens into several micro batches and pipeline their execution one after the other as shown at the bottom of Figure~\ref{fig:mpipe}. Upon completing the first All-to-All for a micro-batch, experts asynchronously execute calculation while simultaneously starting to receive another mini-batch. Then, the second All-to-All operation starts as soon as the calculation is finished.
Furthermore, there is no dependency among operations of different partitions. Thus, we schedule $S$ and $R$ to be executed in the alternative as shown in Figure~\ref{fig:sharded-pipe} for the better locality of memory accesses. The workflow ``communication $\rightarrow$ computation $\rightarrow$ communication" is symmetric in the backward pass.

\begin{figure}[t]
\centering
\subfigure[FasterMoe's fashion] {
 \label{fig:fastermoe}     
\includegraphics[width=0.98\columnwidth]{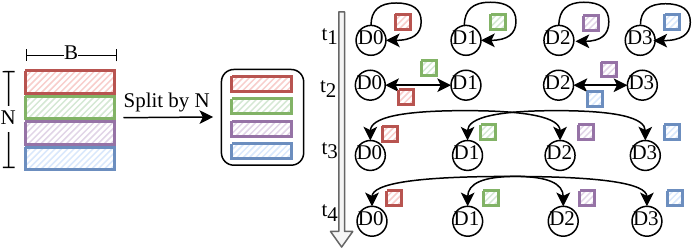}  
}     
\subfigure[MPipeMoe's fashion] { 
\label{fig:mpipe-topo}     
\includegraphics[width=0.98\columnwidth]{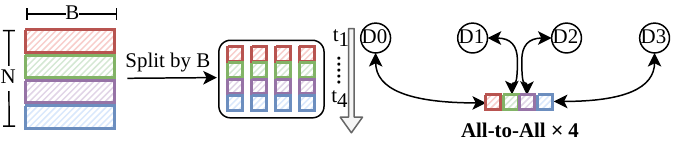}     
}  
\caption{
Comparison between FasterMoE and our methods.
}
\label{fig:way-of-split}
\end{figure}

\textbf{Comparison with FasterMoE in Pipeline Parallelism}.
FasterMoE\cite{fastermoe} also adopts pipeline parallelism to improve the efficiency of MoE training. Different from FasterMoE, we apply a distinguishing method to split the batch data and propose a new optimization solution for communication. As shown in Figure~\ref{fig:way-of-split}, the shape of tensor $T_I$ is $(N, B)$, the first dimension is the number of devices while the second is the batch size of tokens. 
Each row of the tensor is assigned to the device, which is indicated in a different color in the figure. There exist two methods for splitting $T_I$ into multiple partitions. 
The first method splits $T_I$ along the first dimension. The All-to-All operation is partitioned into several point-to-point communications among workers for each partition as shown in Figure~\ref{fig:fastermoe}. 
The second method splits $T_I$ along the second dimension as shown in Figure~\ref{fig:mpipe-topo}. The original All-to-All is split into a few fine-grained ones, each for one partition.
FasterMoE adopts the former method, which has two disadvantages. First, the All-to-All communication is broken down into multiple point-to-point communications, making it infeasible to take advantage of optimizations offered by NCCL. Second, in the phase of communication, if the network bandwidth is heterogeneous among workers, the synchronization procedure causes a waste of resources for those workers with higher bandwidth. As a result, MPipeMoE adopts the latter method for better performance.

\IncMargin{1em}\begin{algorithm}[htb]\label{al:gran-search}
    \caption{Adaptive Pipeline Granularity Search}
    \SetAlgoNoLine 
    \LinesNumbered 
    \KwIn{the batch size of tokens $B$}
    \KwOut{the number of partitions $n$}
    \textbf{global:} $\mathcal{S} = \{\}$ \; 
    \textbf{global:} $cache\_table = \{\}$ \; 
    
    \If{$B\ in\ cache\_table$}{
        return $ cache\_table[B]$ \;
    }
    $(\mathcal{R}_n,n) = find(\mathcal{S}, B)$ \;
    \If{$n == -1$}{
        $n = searchBestGran(B)$ \;
        $(\mathcal{R}_n, n) = find(\mathcal{S}, B)$ \;
        \eIf{$\mathcal{R}_n == \varnothing $}{
            $\mathcal{R}_n = range(B, B)$ \;
            $insert(\mathcal{S}, (\mathcal{R}_n, n))$ \;
        }{
            $\mathcal{R}_n = range(min(B, B_n^{lower}), max(B, B_n^{upper}))$ \;
        }
    }
    $ cache\_table[B] = n$ \;
    return n\;

\end{algorithm}\DecMargin{1em}  

\subsection{Adaptive Pipelining Granularity Configuration} 
\label{sec:dynamic-pipe}

The effectiveness of pipeline parallelism is largely determined by the granularity of pipeline, which is determined by the number of partitions $n$. A coarse-grained granularity fails to take the benefit of pipeline because $S$, $C$, and $R$ cannot be fully overlapped. On the other hand, a very fine-grained granularity could lead to GPU under-utilization. Therefore, it is necessary to configure for the optimal $n$ at runtime to take full advantage of pipeline parallelism.

We consider the training process of MoE models, in which the batch size of tokens is split into $n$ partitions. The micro-batch size equals $B/n$. It requires running dozens of iterations to search for the optimal configuration of $n$ by calling the method $searchBestGran(B)$. Although the cost can be amortized by epochs, unfortunately, $B$ is dynamic and span a wide range in MoE training~\cite{tutel}. Thus, it is time consuming to search for the optimal $n$ for every value of $B$.

In order to reduce the searching space, we propose Algorithm~\ref{al:gran-search} based on an  intuitive hypothesis: $n$ is monotonically increasing as $B$ increases. 
As a result, the whole value domain of $B$ can be a set of disjoint ranges $\{\mathcal{R}_n\}(\mathcal{R}_n=range(B_n^{lower}, B_n^{upper}))$, which is a one-to-one mapping to $n$. We denote the set of pairs $(n,\mathcal{R}_n)$ as $\mathcal{S}$. Given the batch size of tokens $B$, the optimal $n$ can be looked up by finding a pair $(n,\mathcal{R}_n)$ that satisfies $B\in\mathcal{R}_n$ in $\mathcal{S}$, which is shown in line 6. 
If not found, $searchBestGran(B)$ is called to search for the optimal configuration $n$ by trials, i.e., lines 7-8. If $n$ is not in $\mathcal{S}$, a new pair $(n, \mathcal{R}_n=(B,B))$ is inserted into $\mathcal{R}s$, i.e., lines 9-12. Otherwise, we merge $B$ into range $\mathcal{R}_n$, as shown in lines 13-14. To eliminate the overhead of $find(B)$ method, we build a hash table to cache the best strategy for each specific $B$, which is illustrated in lines 3-5.
We implement the set $\mathcal{R}s$ based on the binary-search-tree algorithm. The complexity of $find(B)$ and $insert(n, B)$ are both $O(log(n))$.

\subsection{Memory Reusing} \label{sec:design-shraded}

\begin{figure}[htbp] \label{fig:opportunity}
\centering
\scalebox{0.35}{
\centerline{\includegraphics{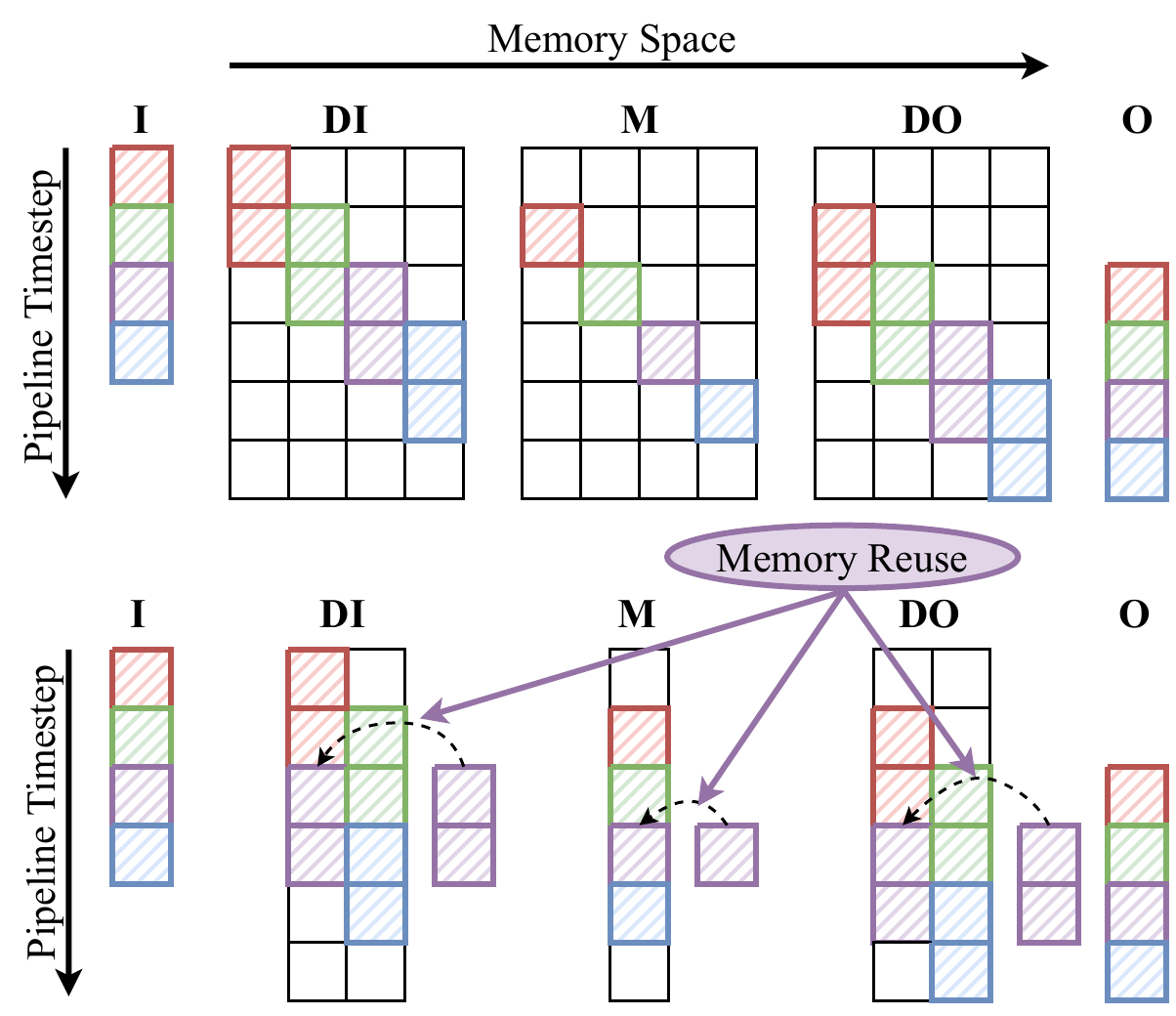}}
}
\caption{The illustration of memory reusing. The top figure demonstrates ``memory bubbles'' in pipeline parallelism and the bottom one shows the compressed memory by memory reusing.}
\label{fig:opportunity}
\end{figure}

Tensors $T_{DI}$, $T_{M}$, and $T_{DO}$ are split into $n$ partitions in pipeline parallelism. Different partitions of tensors are activated at different times, resulting in ``memory bubbles'' as shown at the top of Figure~\ref{fig:opportunity}. The same operation on different partitions is pipelined into a single stream and executed in sequence. We demonstrate that the input or output tensors of these operations can be shared among partitions to reduce memory redundancy. For example, the \textit{i-th} partition of tensor $T_{M}$ is activated for computation at time $t$ and the \textit{(i+1)-th} partition is activated at time $t+1$. Thus we just can allocate one buffer memory to store partitions of $T_{M}$ in turn. In this way, the required memory is reduced from $m$ to $\frac{m}{n}$, where $m$ is the original memory requirement. Similarly for $T_{DI}$ and $T_{DO}$, it requires two buffers for communication and computation as shown at the bottom case of Figure~\ref{fig:opportunity}.

The memory reusing method is applicable for temporary buffers. The peak memory requirement of temporary buffers equals that of activations in pipeline parallelism, thus we can obtain $\mathcal{M}_{buf}^{pipe}$ in Equation~\ref{eq:mem-pipe}. 
With memory reusing, the corresponding reduced memory $\Delta\mathcal{M}_{buf}$ equals $\Delta\mathcal{M}_{act}$, which is presented in Equation~\ref{eq:reduced-mem}.
Finally, we can obtain the memory saving ratio $\phi$ as formulated in Equation~\ref{eq:estimate-mem-save}.
\begin{gather}
\mathcal{M}_{buf}^{pipe} = \mathcal{M}_{act}^{pipe} = 4* B* M+B* H \label{eq:mem-pipe} \\
\Delta \mathcal{M}_{buf} = \Delta \mathcal{M}_{act} = B*(2M*\frac{n-2}{n}+H*\frac{n-1}{n})	\label{eq:reduced-mem} \\
\begin{aligned} \label{eq:estimate-mem-save}
    \phi  &= \frac{\Delta\mathcal{M}_{act}+\Delta\mathcal{M}_{buf}}{\mathcal{M}_{ms}+\mathcal{M}_{act}^{pipe}+\mathcal{M}_{buf}^{pipe}} 
\end{aligned}
\end{gather}

After eliminating memory redundancy, tensors $T_{DI}, T_{M}$ are overridden by other partitions. However, these tensors are required for computing the gradients in the backward pass. To restore tensors $T_{DI}, T_{M}$, we consider two methods as follows.

\begin{figure}[htbp]
\centering
\subfigure[pipeline parallelism] {
 \label{fig:sharded-pipe}     
\includegraphics[width=0.93\columnwidth]{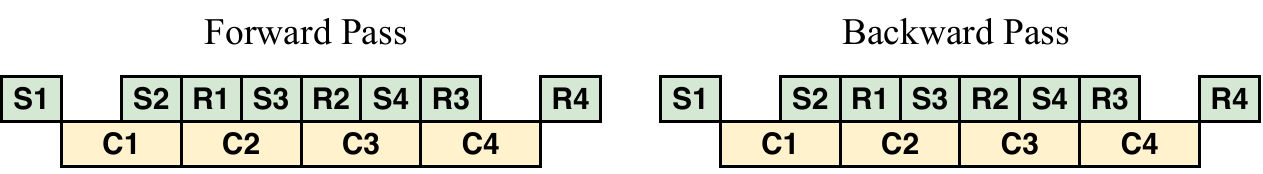}  
}     
\subfigure[S1] { 
\label{fig:sharded-s1}     
\includegraphics[width=0.93\columnwidth]{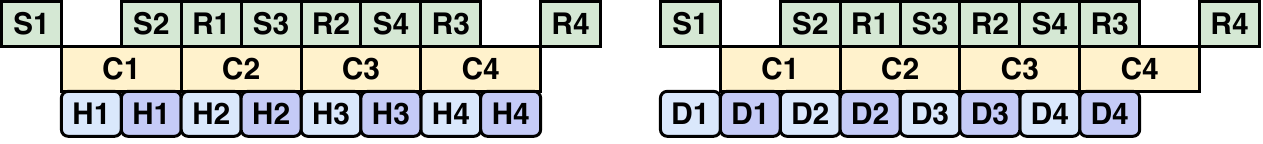}     
}
\qquad
\subfigure[S2] {
 \label{fig:sharded-s2}     
\includegraphics[width=0.93\columnwidth]{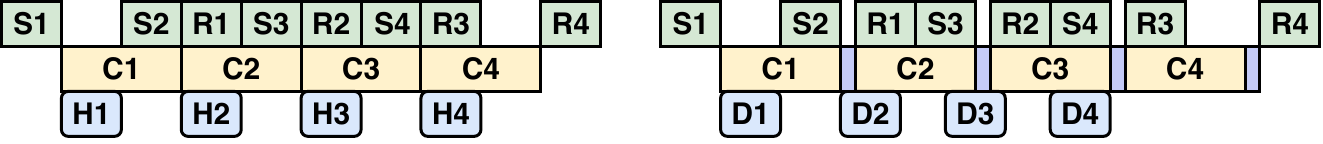}  
}     
\subfigure[S3] { 
\label{fig:sharded-s3}     
\includegraphics[width=0.93\columnwidth]{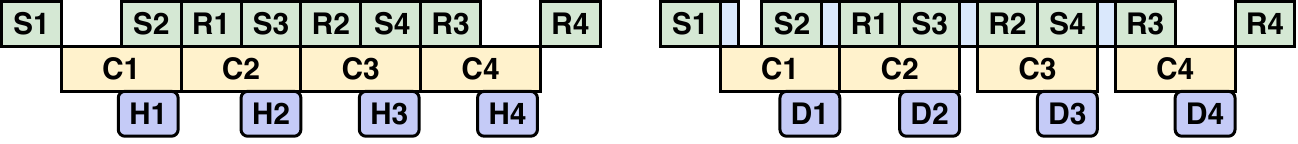}     
}
\subfigure[S4] {
 \label{fig:sharded-s4}     
\includegraphics[width=0.93\columnwidth]{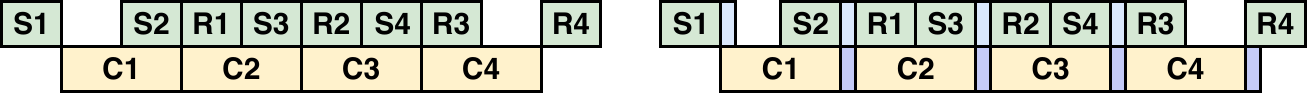}  
}     
\caption{
    The timeline of pipeline parallelism and memory reusing. $H, D$ represent the host-to-device and device-to-host memory copies, respectively. 
}
\label{fig:sharded}
\end{figure}

\begin{itemize}
    \item Data offloading. Leveraging the fact that modern GPUs support overlapping computations and data transfers, we can swap data back to the CPU while computing. In the backward pass, data can be prefetched to the GPU memory in advance. 
    \item Communication and Re-computation.  
    Tensor $T_{DI}$ can be transferred again from tensor $T_I$. And $T_M$ can be re-computed from $T_{DI}$. Ideally, the additional cost of re-computation can be mitigated if communication is the bottleneck and vice versa.
    
\end{itemize}

As a result, we have four memory reusing strategies, i.e., $S1$, $S2$, $S3$, and $S4$ for MoE training, which are illustrated in Figure~\ref{fig:sharded-s1}-\ref{fig:sharded-s4}. These strategies distinguish in adopting different methods to restore $T_{DI}$ and $T_{M}$ in the backward pass.
Because there is no dependency among operations of different partitions, we schedule $S$ and $R$ in Figure~\ref{fig:mpipe} to be executed in an alternative manner for the better locality of memory accesses.
Compared with the timeline of the pipeline without a memory reusing strategy as shown in Figure~\ref{fig:sharded-pipe}, $S1$, $S2$, and $S3$ require another CUDA stream to perform memory copy operations in parallel with computation and communication. 
Specifically, device-to-host and host-to-device memory copy operations are involved in the forward pass and the backward pass, respectively. 
In $S2$ and $S4$, additional communication operations are introduced to restore $T_{DI}$ in the backward pass. Additional computation operations are also required to restore $T_M$ in $S3$ and $S4$.


\begin{table}[t]
\caption{Different Strategies for Memory Reusing} \label{tab:strageties}
\centering
\scalebox{1}{
\begin{tabular}{clllll} 
\toprule
strategy  &  $T_{DI}$ & $T_{M}$ & $\mu$ & $\eta$ & $\mathcal{Q}_{fw}, \mathcal{Q}_{bw}$  \\
 \midrule
none & - & -  & $\mu_{comp}$ & - & [2,2,0],[4,2,0] \\
S1 & offload & offload & $\mu_{all}$ & $\eta_{all}$ & [2,2,5],[4,2,5]  \\
S2 & comm. & offload   & $\mu_{all}$ & $\eta_{all}$ & [2,2,4],[4,3,4]\\
S3 & offload & recompute &  $\mu_{all}$ & $\eta_{all}$  & [2,2,1],[5,2,1] \\
S4 & comm. & recompute  & $\mu_{comp}$ & - & [2,2,0],[5,3,0] \\
\bottomrule
\end{tabular}}
\label{tab:strageys}
\end{table}

\subsection{Performance Model on Memory Reusing Strategies}\label{sec:design-dyn-strategy}

In Section~\ref{sec:speed-profile}, we validate the feasibility of pipeline parallelism and denote the speed of computation, communication, and memory copy as $\sigma_xW_{comp}$, $\mu_xW_{comm}$, and  $\eta_xW_{mem}$, in which $x$ refers to the interference stream.
For simplicity, we define $v_0=[v_{0, comp}, v_{0,comm}, v_{0,mem}]$ as the amount of different type of operations in Equations~\ref{eq:t0-comp} to \ref{eq:t0-mem}, where $H, M$ are defined in Table~\ref{tab:notations}.
Specifically, $v_{0, comp}$ and $v_{0, comm}$ are the amount of floating-point operations and All-to-All collective data volumes in MoE, respectively. 
$v_{0,mem}$ represents the amount of data volumes produced by moving tensor $T_{DI}$ between the device and host. Because $H=4* M$ in most MoE models, copying tensor $T_{M}$ requires four times more data than that of $v_{0,mem}$.

To quantify the workload on three streams, we define $\mathcal{Q}=[q_1, q_2, q_3]$ to represent the actual amount of relative operations. 
For instance, if not performing any memory reusing strategy, i.e., $\mathcal{Q}_{fw}=[2,2,0]$, there exists two GeMM operations and two All-to-All operations in the forward pass. And similarly, we can obtain $\mathcal{Q}_{fw}$ and $\mathcal{Q}_{bw}$ of four memory reusing strategies in Table~\ref{tab:strageys}.
\begin{gather}
    v_{0,comp} = b* H* M \label{eq:t0-comp} \\
    v_{0,comm} = b* M \label{eq:t0-comm} \\
    v_{0,mem} = b* M \label{eq:t0-mem} 
\end{gather}

The execution time of a specific stream equals the total amount of operations divided by the processing speed. For instance, the computation time is $\frac{q_1 v_{0,comp}}{\sigma W_{comp}}$.
Because different CUDA streams execute different tasks in parallel, the execution time of the end-to-end pipeline is determined by the slowest stream. We formulate the overall execution time $\mathcal{Q}$ in Equation~\ref{eq:cost-general}, which is determined by $\mathcal{Q}$, $\mu$, and $\eta$. Because the floating-point operations per second are stable as stated in Section~\ref{sec:speed-profile}, $\alpha$ and $\beta$ are nearly constant. 

\begin{gather}
\begin{aligned}
    \mathcal{C} &= max(\frac{q_1 v_{0,comp}}{\sigma W_{comp}}, \frac{q_2 v_{0,comm}}{\mu W_{comm}}, \frac{q_3 v_{0,mem}}{\mu W_{mem}}) \\
    &\approx \frac{1}{W_{comp}}max(q_1, q_2\alpha/\mu, q_3\beta/\eta) \\
    &= \frac{1}{W_{comp}} max(\mathcal{Q}\cdot[1, 1/\mu, 1/\eta] \cdot [1, \alpha, \beta ])  \label{eq:cost-general}
\end{aligned} 
\end{gather}
\begin{equation*}
 \text{in which}  \ \ \ \alpha = \frac{W_{comp}}{W_{comm}} , \ \ \ \ \ 
    \beta = \frac{W_{comp}}{W_{mem}}
    \label{eq:alpha-beta}     
\end{equation*}

Table~\ref{tab:strageties} summarizes the characteristics of four strategies, i.e., $\mu$, $\eta$, $\mathcal{Q}_{fw}$, and $\mathcal{Q}_{bw}$.
We obtain the cost $\mathcal{C}$ for all strategies based on Equation~\ref{eq:cost-general}, from which the one with the lowest cost is chosen as the optimal memory reusing strategy. Generally, strategies $S1$ and $S2$ introduce more memory copy operations, which tend to be I/O bound. In contrast, strategies $S3$ and $S4$ tend to be compute-bound.

\section{Implementation}\label{sec:impl}

MPipeMoE is an end-to-end MoE training library implemented on top of torch 1.9.0 with CUDA 11.1~\footnote{https://github.com/whuzhangzheng/MPipeMoE}.  A few key components and functionalities are implemented as follows.

\subsection{Gating network and Experts} 
The gating network routes tokens to experts based on top-$k$ algorithm. In this paper, we set $k$ to 1. Increasing $k$ is an equivalence of increasing $B$ in the perspective of system performance.  We implement a feed-forward network as the default expert, which is applicable for most transformer models.

\subsection{Expert Parallelism} MPipeMoE distributes experts across different GPUs using the expert parallelism while running the remaining parts of the MoE model in data-parallel scheme. NCCL All-to-All collective operator is adopted to dispatch and collect tokens among GPUs.

\subsection{Usability} It is easy for users to benefit from optimizations introduced by MPipeMoE using Python API, including adaptive pipeline parallelism and memory reusing, etc. As shown in the following code snippet, adding parameters such as \textit{pipeline} and \textit{memory\_reuse} can bring the end-to-end speedup.
\begin{lstlisting} 
import pmoe
moe_layer = pmoe.MoELayer(d_model=1024, 
    d_hidden=4096, top_k=1, 
    num_experts=64, pipeline=True, 
    memory_reuse=True)
\end{lstlisting}








\section{EVALUATION} \label{sec:exp}

\subsection{Experimental Setup} \label{sec:exp-setup}

\subsubsection{Physical cluster} Experimental evaluation is performed on a physical cluster consisting of 8 \texttt{NVIDIA DGX A100} servers. Each node is equipped with 8 NVIDIA A100 SXM 40GB GPUs and 200 Gbps HDR InfiniBand, backed by 96 $\times$ 2nd-generation AMD EPYC CPU cores and 1.9 TiB memory. GPUs are connected by the 3-rd generation NVLink and NVSwitch within each machine. GPUs across different machines are connected through a 1,600 Gbps InfiniBand network with adaptive routing.

\subsubsection{Models and configurations} The significant discrepancy of the MoE layer between different models lies in the size of experts, which is determined by $M$ and $H$, and the number of tokens that is denoted as $B$. Here, we aim to validate the efficiency of the proposed methods on different expert sizes as well as batch sizes.  We configure the different expert sizes of FFN from BERT~\cite{bert} and GPT-3~\cite{gpt3} as shown in Table~\ref{tab: settings}, in which $d_{model}$ refers to the dimension of token embedding and $d_{hidden}$ refers to the hidden dimension of FFN layer of models, respectively. Note that this paper focuses on the MoE structure and related performance. We create a dummy dataset by generating random tokens as input to different models. For all experiments, we adopt Adam~\cite{adam} as the optimizer. We evaluate the efficiency of MPipeMoE in terms of the average training time and the peak memory footprint.

\subsection{Methodology} \label{sec:exp-methodology}

\begin{table}[t]
\caption{Specifications of MoE layers}
\centering
\scalebox{1}{
\begin{tabular}{lccc} 
\toprule
 Model Name & $d_{model}$ & $d_{hidden}$ & \#experts\\
 \midrule
MoE-GPT3-S & 768  & 3072 & 64\\
MoE-GPT3-XL & 2048 & 8192 & 64\\
MoE-BERT-L & 1024 & 4096  & 64 \\
\bottomrule
\end{tabular}}
\label{tab: settings}
\end{table}

To demonstrate the performance gain and memory efficiency, we compare MPipeMoE against the state-of-art system \textit{FasterMoE}~\cite{fastermoe}, which implements dynamic shadowing and pipeline parallelism in MoE training. We also choose \textit{FastMoE} as another competitor, which implements the primitive expert parallelism without pipeline parallelism.

We implement \textit{MPipeMoE} and its variant \textit{PipeMoE} to demonstrate the advantages of adaptive pipeline parallelism and memory efficiency. \textit{PipeMoE} implements micro-batch size splitting, which also adopts multi CUDA streams to execute computation and communication in parallel. \textit{MPipeMoE} is implemented on top of \textit{PipeMoE}, which adopts adaptive memory reusing strategies to further reduce memory footprint.




\subsection{Overall Speedup} \label{sec:exp-pp}
\begin{figure}[t] 
    \centering
    \scalebox{0.65}{ 
    \includegraphics[width=.75\textwidth,trim=0 0 0 0,clip]{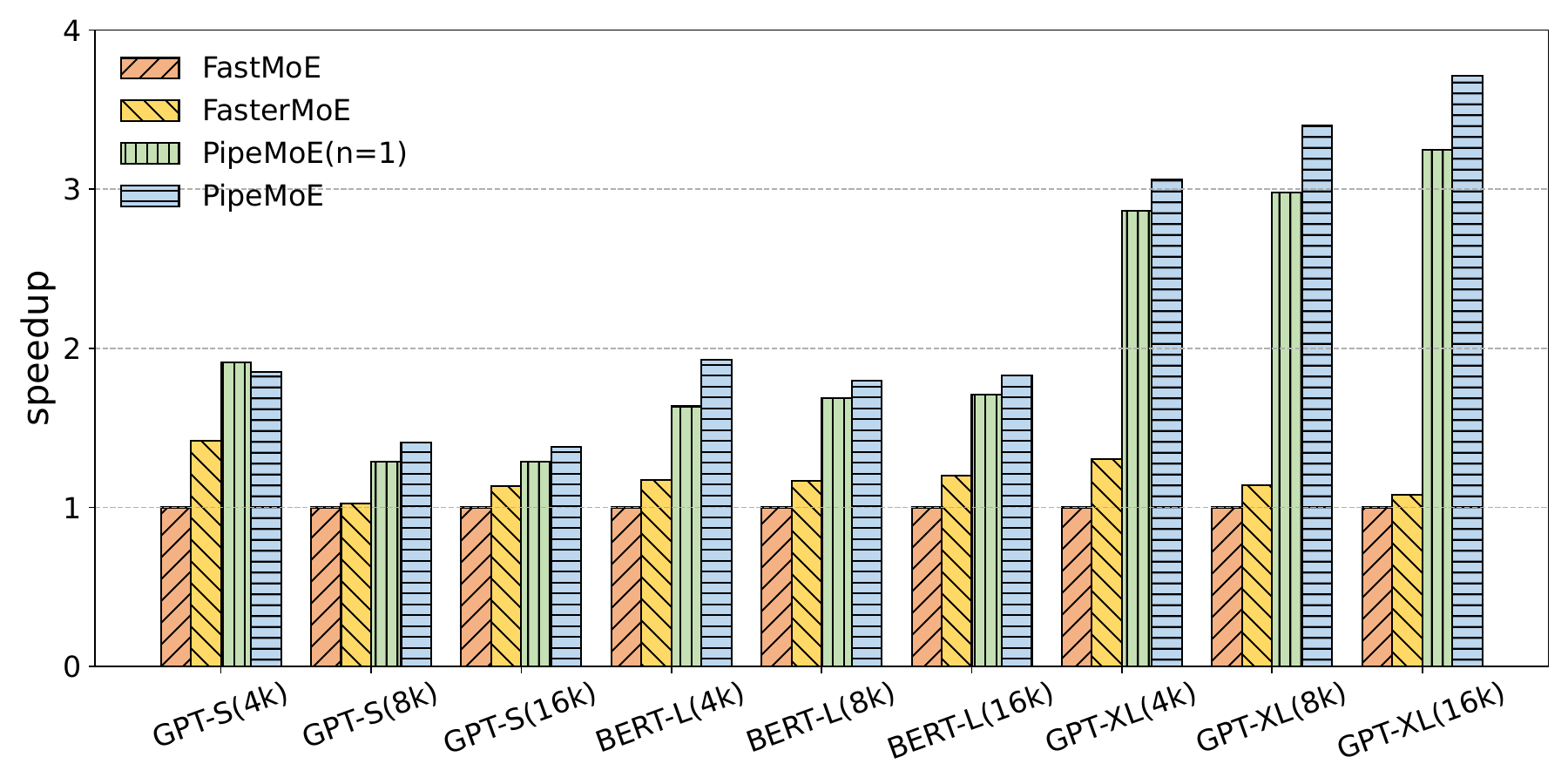}
    }
    \caption{The speedup of different methods in MoE training with the same model setting and batch size of tokens $B$. The format of x-axis is ``model\_name($B$)".}
    \label{fig:pipe-time}
\end{figure}

Figure~\ref{fig:pipe-time} shows the speedup of PipeMoE against FastMoE and FasterMoE in model training. Compared with FasterMoE, PipeMoE achieves an average speedup of 2.26$\times$ on various models and batch sizes. Compared with FastMoE, PipeMoE achieves up to 3.7$\times$ speedup. FasterMoE outperforms FastMoE because of pipeline parallelism and overlapping of  computation and communication. PipeMoE can improve the speedup up to 3.4$\times$ against FasterMoE, largely because of the optimization of the pipeline granularity. PipeMoE also takes advantage of Tensor Core of GPUs to accelerate computation.

To validate the effectiveness of pipeline parallelism, we compare PipeMoE against PipeMoE($n$=1). In PipeMoE($n$=1), the communication and computation are executed in sequence. From the result, we can see that the implementation of pipeline brings benefits to various models with different batch sizes of tokens. The only exception is GPT-S with batch size $B=4k$, which is not a computation-intensive workload. The result indicates that the pipeline cannot benefit the training process that is not compute-bound because the additional kernel launch overhead leads to lower GPU utilization. 



\subsection{Memory Footprint Reduction}  \label{sec:exp-mem}
\begin{figure}[t] 
    \centering
    \scalebox{0.65}{
    \includegraphics[width=.75\textwidth,trim=0 0 0 0,clip]{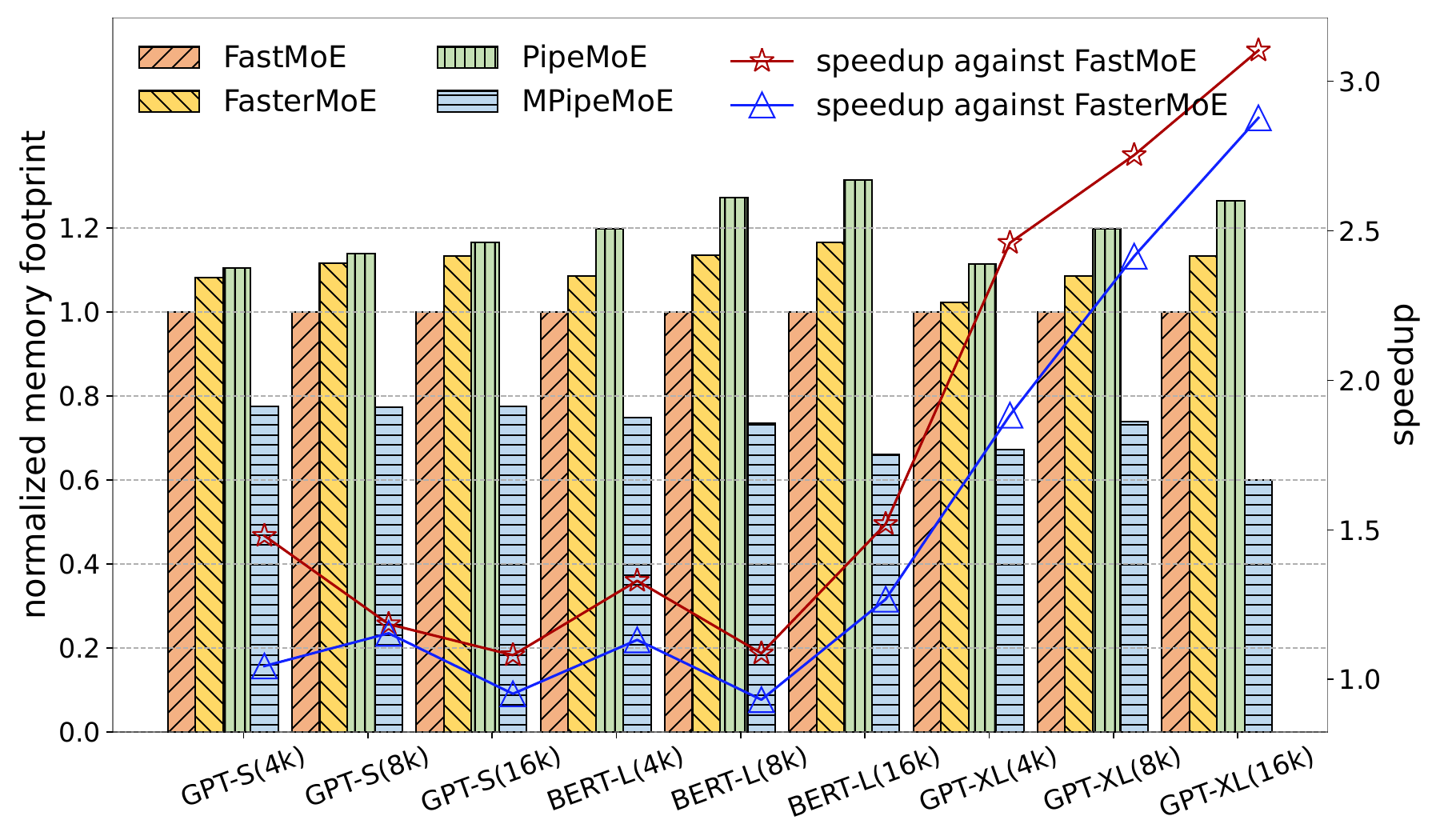}
    }
    \caption{The memory footprint reduction by MPipeMoE. The bars and the left y-axis show the ratio of memory footprint compared to FastMoE. The polyline and the right y-axis show the speedup of MPipeMoE compared to FastMoE and FasterMoE, respectively.}
    \label{fig:memory}
\end{figure}

\begin{figure}[t] 
    \centering
    \scalebox{0.65}{
    \includegraphics[width=.75\textwidth,trim=0 0 0 0,clip]{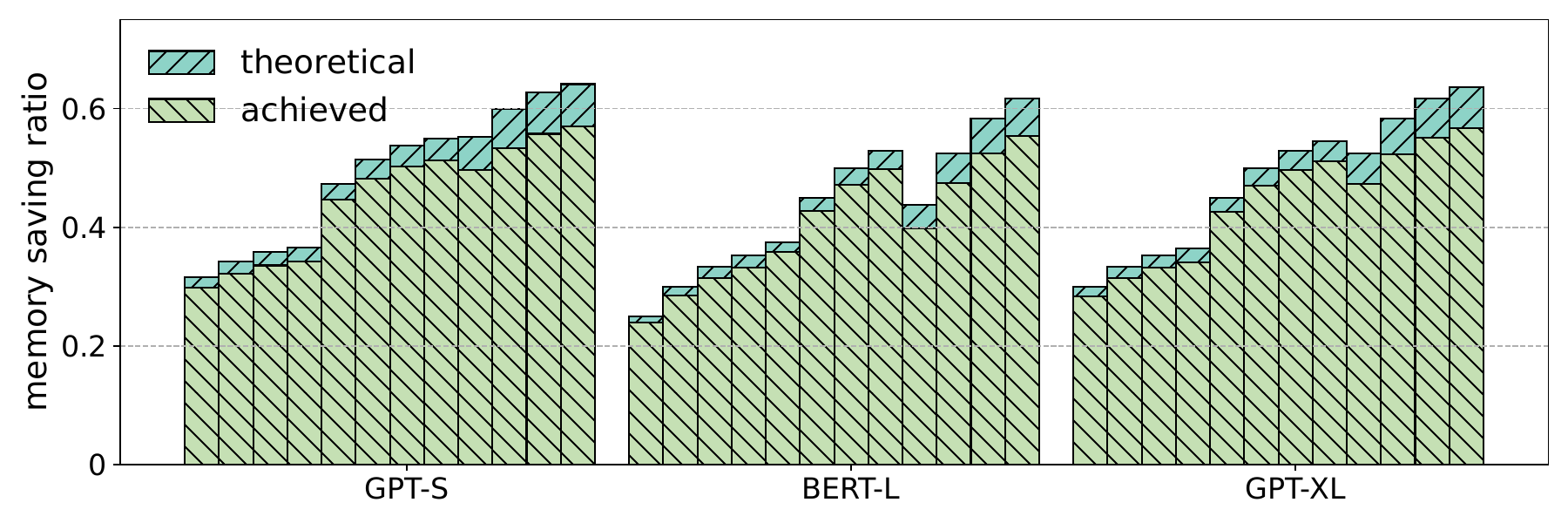}
    }
    \caption{The MPipeMoE achieved memory reduction ratios compared to their theoretical results on three model settings with the varying number of partitions $n$ (2,4,8) and batch sizes (ranging from 4k to 32k).}
    \label{fig:memory-estimate}
\end{figure}

    Figure~\ref{fig:memory} presents the overall memory footprint of the approaches, where the left y-axis represents the memory footprint normalized to that of FastMoE. The result shows that MPipeMoE reduces the memory footprint by an average of 23\% and up to 40\% compared to FastMoE while still can achieve 3.1$\times$ speedup in terms of training time. FasterMoE requires more memory than FastMoE because of the dynamic shadowing and smart scheduling. As a result, MPipeMoE achieves an average memory reduction of 27\% and up to 47\% compared with FasterMoE. Meanwhile, MpipeMoE achieves a speedup up to 2.8$\times$ in terms of the training time. 
    
    In Section~\ref{sec:design-shraded}, Equation~\ref{eq:estimate-mem-save} provides the theoretical bound of memory saving of MPipeMoE. To demonstrate the effectiveness of the analysis, we report the actually achieved memory saving ratio with the bound, which is depicted in Figure~\ref{fig:memory-estimate}. We conduct experiments on three models. We configure the number of partitions $n$ and the batch size of tokens $B$ to different values to validate a wide range of cases. MPipeMoE achieves about 95\% of the theoretical bound. Note that tensors with small sizes such as routing data produced by gating networks are not considered.

\subsection{Performance Breakdown} \label{sec:exp-breakdown}
\begin{figure}[t] 
    \centering
    \scalebox{0.65}
    {
    \includegraphics[width=.75\textwidth,trim=0 0 0 0,clip]{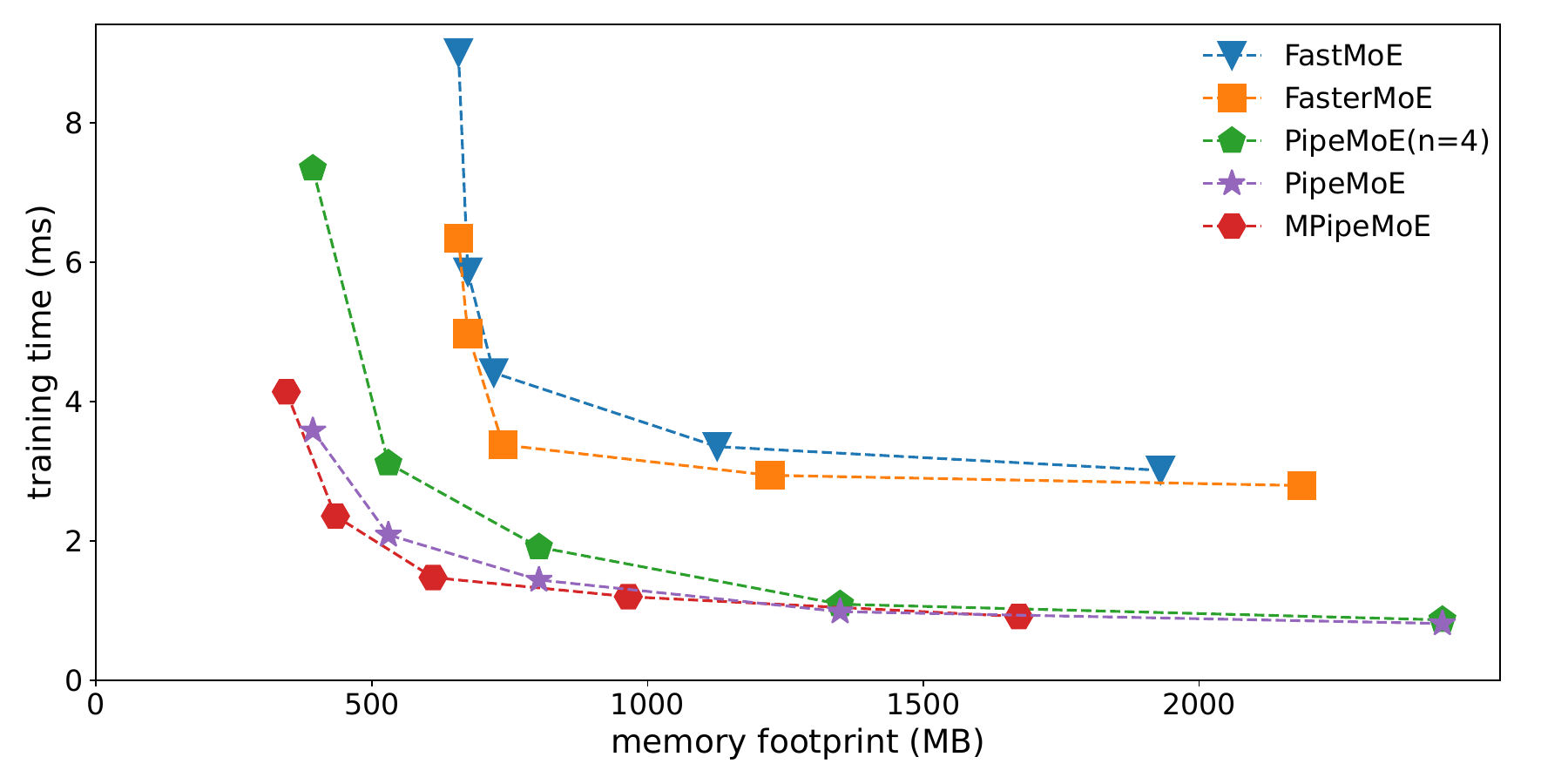}
    }
    \caption{Overall performance breakdown of MPipeMoE on GPT-XL model.}
    \label{fig:time-mem}
\end{figure}

To better understand the performance breakdown of MPipeMoE, we reveal the performance of the different methods in \textit{memory-time} coordinates, in which the x-axis represents the memory footprint and the y-axis represents the training time. As shown in Figure~\ref{fig:time-mem}, the one closer to the origin point illustrates better overall performance. MPipeMoE significantly outperforms FasterMoE and FastMoE. 
 PipeMoE(n=4) reduces the training time because of a higher GPU throughput. PipeMoE outperforms PipeMoE(n=4) by configuring the optimal pipeline  granularity at runtime. 
 MPipeMoE achieves best memory efficiency by reusing memory partitions. The higher GPU utilization makes it possible to increase the batch size with the limited device memory space.



\subsection{Effectiveness of Granularity Configurations} \label{sec:exp-dyngran}
\begin{figure}[t] 
    \centering
    \scalebox{0.65}{
    \includegraphics[width=.75\textwidth,trim=5 0 0 0,clip]{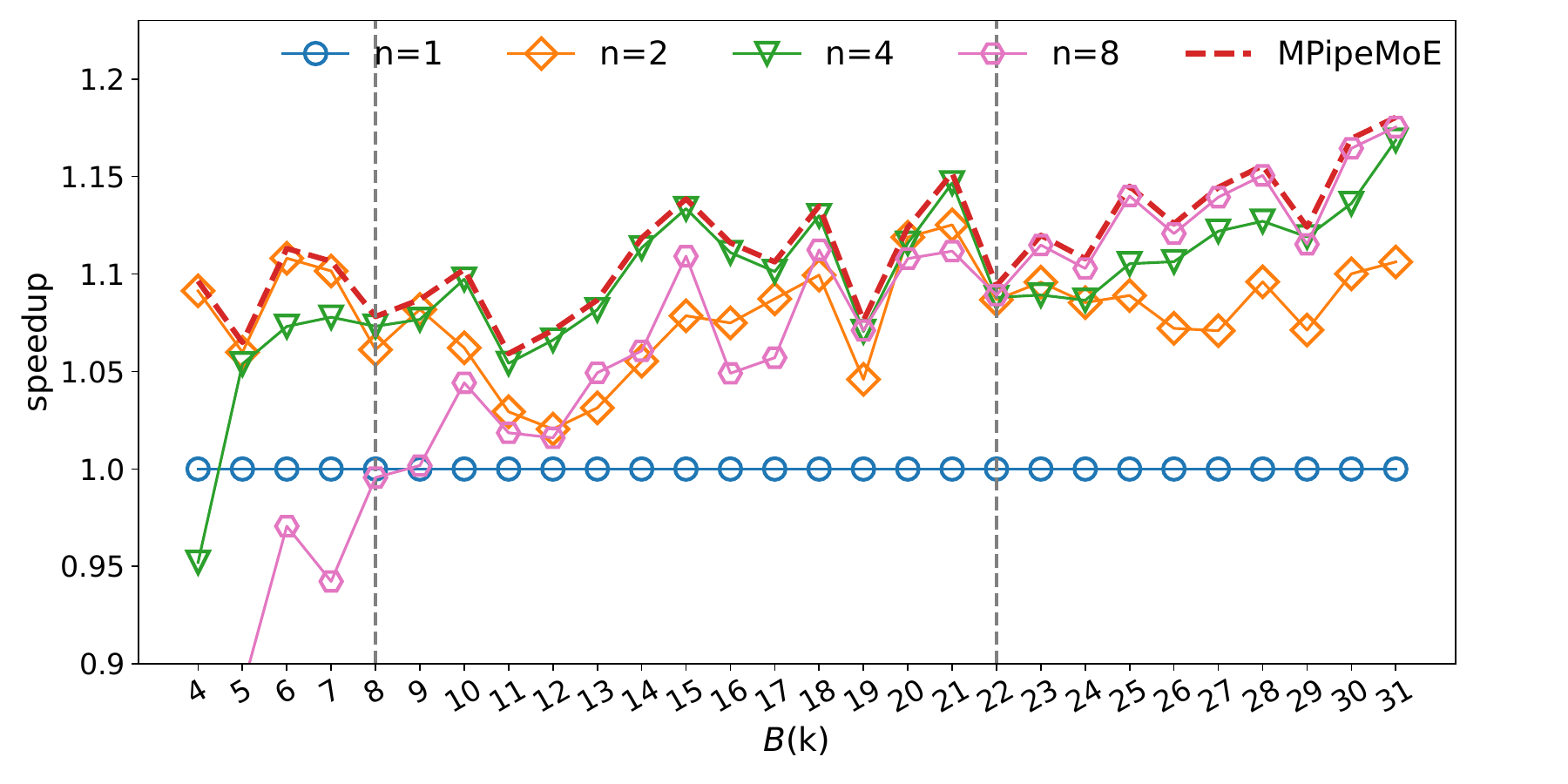}
    }
    \caption{The effects of pipeline parallelism on various pipeline granularity. The dashed line represents the adaptive granularity selected by the configuration algorithm. The x-axis represents various $B$ values.}
    \label{fig:pipe-adaptive}
\end{figure}

    We illustrate the effectiveness of the adaptive pipeline granularity configuration of MPipeMoE, which is based on a hypothesis that $n$ is monotonically increasing as $B$ increases. We compare the performance due to different pipeline granularity with various batch sizes of tokens on model GPT-XL. Figure~\ref{fig:pipe-adaptive} shows that when the batch size is smaller than 8k, $n=2$ is the best option. When the batch size is increased to 8k-22k, $n=4$ ensures the best performance. $n=8$ is the optimal configuration if the batch size is larger than 22k. MPipeMoE, which is denoted as a dashed line, performs the best in 
   all situations. The results validate its effectiveness.

\subsection{Overhead of Memory Reusing} \label{sec:exp-stategy-select}

\begin{figure}[t] 
    \centering
    \scalebox{0.65}{
    \includegraphics[width=.75\textwidth,trim=0 0 0 0,clip]{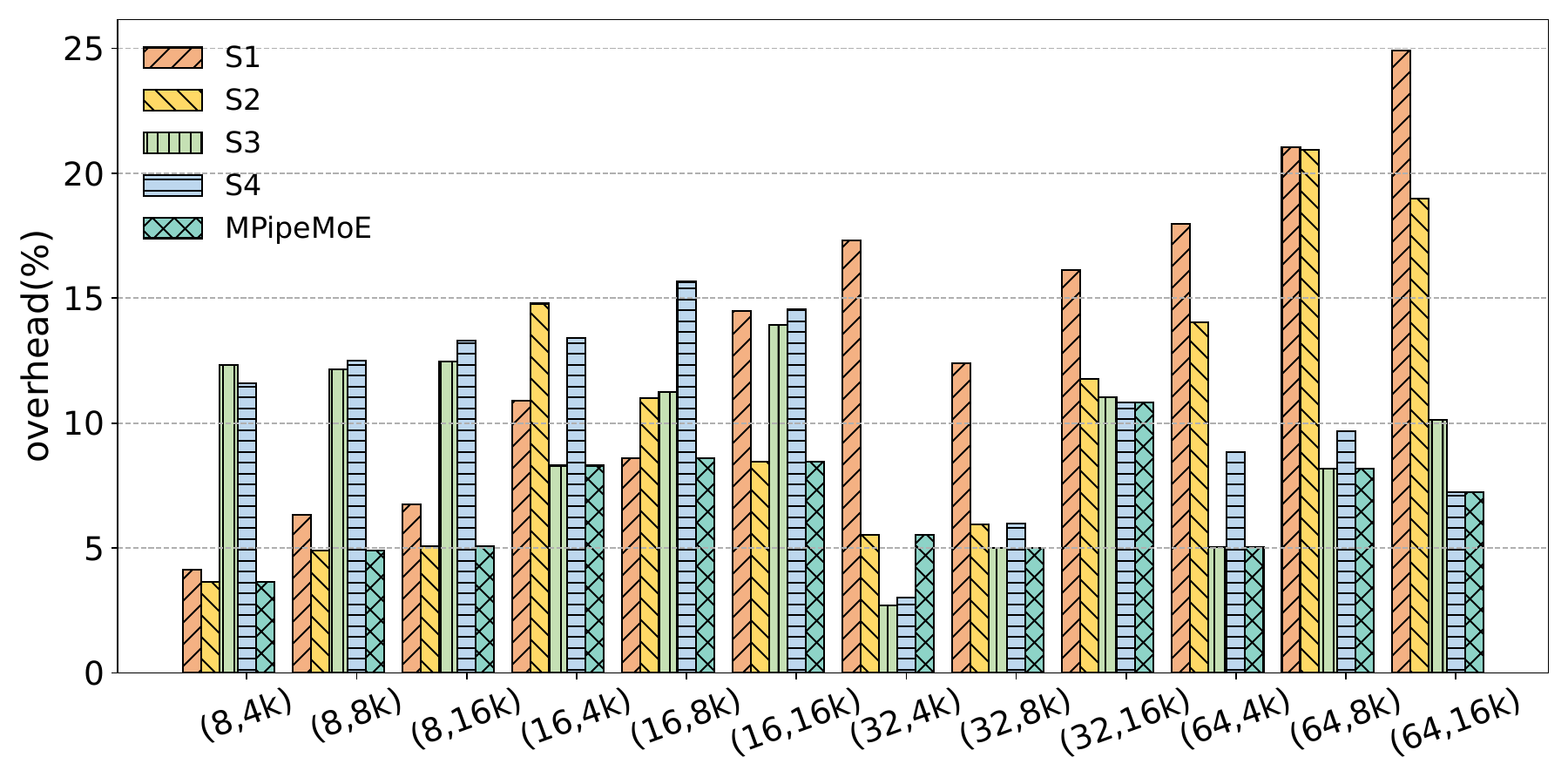}
    }
    \caption{The overhead of memory reusing strategies and the effectiveness of the strategy selection method in MPipeMoE. The ticks of the x-axis represent different numbers of GPUs $N$ and the batch size of tokens $B$ in format $(N, B)$.}
    \label{fig:dynselect}
\end{figure}

    
    In terms of speedup, MpipeMoE is indeed second to PipeMoE because MpipeMoE achieves memory efficiency at the same time, which however incurs non-trivial overhead. MpipeMoE features four memory reusing strategies, i.e., $S1$, $S2$, $S3$, and $S4$ defined in Table~\ref{tab:strageties}, which resort to re-computation/communication and CPU offloading to restore activation tensors in the backward pass. For overhead analysis of the strategies, we conduct experiments with different numbers of GPUs $N$ and various batch sizes of tokens $B$. Figure~\ref{fig:dynselect} presents the results, from which we can observe that:
    \begin{itemize}
        \item  $S1$ and $S2$ perform better when $N$ is small, e.g., 8, but worse with a larger $N$, e.g., 64. $S1$ and $S2$ introduce additional memory copy operations while $S2$ introduces additional communication operations. 
        With the increasing number of workers, the cost of communication also increases, which results in the worse performance for $S2$ due to the competition on the memory bandwidth between memory copy and communication.
        \item Both $S3$ and $S4$ introduce additional computational costs, which perform worse if the workload is computation-bound, i.e., $N=8$. 
        \item $S4$ performs better than $S2$ if $N$ equals 32 or 64, in which communication is the bottleneck because memory copy over PCIe in $S2$ slows down communication operations.
        \item There is not much performance variation with the varying batch sizes, indicating that the batch size is not sensitive to the configuration of strategy. 
    \end{itemize}

    Based on these observations, we can conclude that there does not exist a single memory reusing strategy which can ensure the best performance under all situations. MPipeMoE builds a performance model based on Equation~\ref{eq:cost-general} to decide the optimal strategy considering both the hardware configurations and runtime characteristics.
    



\section{Related Work}\label{sec:related}

\textbf{Mixture-of-Experts (MoE)}.
Several techniques have been proposed to improve the training efficiency of MoE models. Gating Dropout\cite{gating-drop} allows tokens to ignore the gating network and keeps the input at the local machines, reducing the cross-machine communication. DeepSpeed MoE\cite{ds-moe} proposes the hierarchical All-to-All and implements custom CUDA kernels to scale expert parallelism out to many devices as the latency increases linearly with the increase in devices. FasterMoE~\cite{fastermoe} designs a congestion-avoiding expert selection strategy that relieves network congestion to achieve lower training latency. Z-code multilingual Multitask MoE model~\cite{zcodem3} proposes the Zero Redundancy Optimizer to reduce memory footprint.

\textbf{Data, Model, Pipeline, and Expert Parallelism}.
Parallelization is a key strategy for training large models at scale. For a model that fits in the device memory for training, data parallelism (DP) is used to scale training out to multiple devices. In DP, model parameters are replicated on each device. At each step, mini-batch data is divided evenly across all the data parallel processes, such that each process executes the forward pass and backward pass on a different subset of data samples, and uses averaged gradients across processes to update the model locally. To support training giant models, model parallelism (MP)~\cite{mesh-tf, megatron} and pipeline parallelism (PP)~\cite{mesh-tf}, Pipedream~\cite{pipedream} splits the model among processes in either vertical or horizontal ways. Expert parallelism~\cite{switch} is another form of model parallelism targeting expert parameters of MoE models. In expert parallelism, different experts are placed on different devices and executed in parallel. When experts reside on different GPU devices, explicit communication using the All-to-All primitive is required.



\section{Conclusion} \label{sec:conclude}

MoE is a promising technology to improving model quality by scaling the neural network to an extra-scale. In this paper, we consider high performance and memory efficiency of MoE model training in a holistic manner. Toward this end, we design adaptive pipeline parallelism with online granularity configuration.
Second, we analyze the memory footprint breakdown of MoE training and propose efficient memory reusing strategies to reduce memory requirements by eliminating memory redundancies. What is more, we develop an adaptive selection component to decide whether to offload or recompute the required tensors, which considers both the hardware capacities and model characteristics at runtime.
We implement and integrate these features into MPipeMoE library and perform extensive evaluations. The results show that MPipeMoE achieves 2.8$\times$ speedup and reduces memory footprint by up to 47\% compared to FasterMoE.

\section*{Acknowledgment}

Dazhao Cheng is the corresponding author. The project is supported by the Special Fund of Hubei Luojia Laboratory.

We promise that we have not used any AI generation tools in our paper.


\bibliography{bibtex/main}
\bibliographystyle{bibtex/IEEEtran.bst}

\vspace{12pt}

\end{document}